\documentclass[12pt]{article}
\usepackage{times}
\usepackage{geometry}
\usepackage[utf8]{inputenc}
\usepackage{enumitem,amssymb}
\usepackage{ragged2e}
\usepackage{pifont}
\usepackage[sort,super]{natbib}
\usepackage{epsfig}
\usepackage{wrapfig}
\usepackage{url}
\usepackage[colorlinks=true, allcolors=blue]{hyperref}
\usepackage{color}
\usepackage{amsmath,amsfonts}
\usepackage{graphicx}
\usepackage[table]{xcolor}
\usepackage{multirow}
\usepackage{multicol}
\usepackage{tabu}
\usepackage{array}
\newcolumntype{C}[1]{>{\centering\let\newline\\\arraybackslash\hspace{0pt}}m{#1}}

\begin{document}
\clearpage
\thispagestyle{empty}
\raggedright

\Large Astro2020 Science White Paper \linebreak

\huge Energetic Particles of Cosmic Accelerators II: \linebreak
\ \linebreak
\textit{\Large Active Galactic Nuclei and Gamma-ray Bursts}\linebreak
\normalsize

\noindent \textbf{\underline{Thematic Areas}} \linebreak
\noindent \textit{Primary: Multi-Messenger Astronomy and Astrophysics}\linebreak
\noindent \textit{Secondary: Cosmology and Fundamental Physics}\linebreak



\textbf{Principal Authors:}

Name: Tonia M. Venters\\
Institution: NASA Goddard Space Flight Center\\
Email: \href{mailto:tonia.m.venters@nasa.gov}{tonia.m.venters@nasa.gov}\\
Phone: 301-614-5546\\
\bigskip
Name: Sylvain Guiriec\\ 
Institution: George Washington University\\
Email: \href{mailto:sguiriec@gwu.edu}{sguiriec@gwu.edu}; \href{mailto:sylvain.guiriec@nasa.gov}{sylvain.guiriec@nasa.gov}\\
\bigskip
Name: Amy Y. Lien\\
Institution: NASA GSFC/CRESST/University of Maryland, Baltimore County\\
Email: \href{mailto:amy.y.lien@nasa.gov}{amy.y.lien@nasa.gov}\\

\bigskip
\textbf{Co-authors:}\\
Marco Ajello (Clemson University), Terri J. Brandt (NASA GSFC), Harsha Blumer (West Virginia University), Michael Briggs (University of Alabama, Huntsville), Paolo Coppi (Yale University), Filippo D'Ammando (INAF Istituto di Radioastronomia), Brian Fields (University of Illinois, Urbana-Champaign), Justin Finke (Naval Research Laboratory), Chris~Fryer (LANL), Kenji Hamaguchi (NASA GSFC/CRESST/UMBC), J.~Patrick Harding (LANL), John W. Hewitt (University of North Florida), Brian Humensky (Columbia University), Stanley D. Hunter (NASA GSFC), Hui Li (LANL), Francesco~Longo (University of Trieste/INFN Trieste), Julie McEnery (NASA GSFC), Roopesh Ojha (UMBC/NASA GSFC), Vasiliki~Pavlidou (University of Crete), Maria~Petropoulou (Princeton University), Chanda~Prescod-Weinstein (University of New Hampshire), Bindu Rani (NASA GSFC), Marcos~Santander (University of Alabama), John~A.~Tomsick (UC Berkeley/SSL), Zorawar~Wadiasingh (NASA GSFC), Roland~Walter (University of Geneva)\\
  
\pagebreak
\justifying
\setcounter{page}{1}

\section*{Executive Summary}
\vspace{-2.0ex}
The high-energy universe has revealed that energetic particles are ubiquitous in the cosmos and play a vital role in the cultivation of cosmic environments on all scales. Our pursuit of more than a century to uncover the origins and fate of these cosmic energetic particles has given rise to some of the most interesting and challenging questions in astrophysics. Within our own galaxy, we have seen that energetic particles engage in a complex interplay with the galactic environment and even drive many of its key characteristics (for more information, see the first white paper in this series). On cosmological scales, the energetic particles supplied by the jets of active galactic nuclei (AGN) are an important source of energy for the intracluster and intergalactic media, providing a mechanism for regulating star formation and black hole growth and cultivating galaxy evolution (AGN feedback)\cite{2017NatAs...1E.165H}. Gamma-ray burst (GRB) afterglows encode information about their circumburst environment\cite{Sari:1997qe}, which has implications for massive stellar winds during previous epochs over the stellar lifecycle. As such, GRB afterglows provide a means for studying very high-redshift galaxies since GRBs can be detected even if their host galaxy cannot\cite{Perley:2016bke}. It has even been suggest that GRB could be used to measure cosmological distance scales if they could be shown to be standard candles\cite{Ghirlanda:2006ax,Guiriec:2016bae}.

Though they play a key role in cultivating the cosmological environment and/or enabling our studies of it, there is still much we do not know about AGNs and GRBs, particularly the avenue in which and through which they supply radiation and energetic particles, namely their jets. Despite the enormous progress in particle-in-cell and magnetohydrodynamic simulations, we have yet to pinpoint the processes involved in jet formation and collimation and the conditions under which they can occur. For that matter, we have yet to identify the mechanism(s) through which the jet accelerates energetic particles -- is it the commonly invoked diffusive shock acceleration process or is another mechanism, such as magnetic reconnection, required? Do AGNs and GRBs accelerate hadrons, and if so, do they accelerate them to ultra-high energies and are there high-energy neutrinos associated with them? \textbf{MeV $\gamma$-ray astronomy, enabled by technological advances that will be realized in the coming decade, will provide a unique and indispensable perspective on the persistent mysteries of the energetic universe.} 

This White Paper is the second of a two-part series highlighting the most well-known high-energy cosmic accelerators and contributions that MeV $\gamma$-ray astronomy will bring to understanding their energetic particle phenomena. Specifically, MeV astronomy will:
\begin{itemize}
    \vspace{-1.0ex}
    \item[1.] Determine whether AGNs accelerate CRs to ultra-high energies;
    \vspace{-1.0ex}
    \item[2.] Provide the missing pieces for the physics of the GRB prompt emission;
    \vspace{-1.0ex}
    \item[3.] Measure magnetization in cosmic accelerators and search for acceleration via reconnection.
\end{itemize}

\vspace{-3.0ex}
\section*{Active Galactic Nuclei}
\vspace{-2.0ex}

AGNs are the most luminous persistent sources in the universe and are associated with the most massive compact objects known, supermassive black holes (SMBHs) with masses reaching $\sim 10^{10}$ M$_{\odot}$. Of particular interest are those AGNs that are characterized by strong radio emission and relativistic jets (i.e., radio-loud AGN) because their broadband spectra from radio through $\gamma$-rays are characterized by nonthermal
emission, indicating efficient particle acceleration 
\begin{wrapfigure}{r}{0.625\textwidth}
\begin{center}
\vspace{-3.ex}
\resizebox{4.in}{!}
{\includegraphics[trim = 85mm 70mm 92mm 75mm, clip]{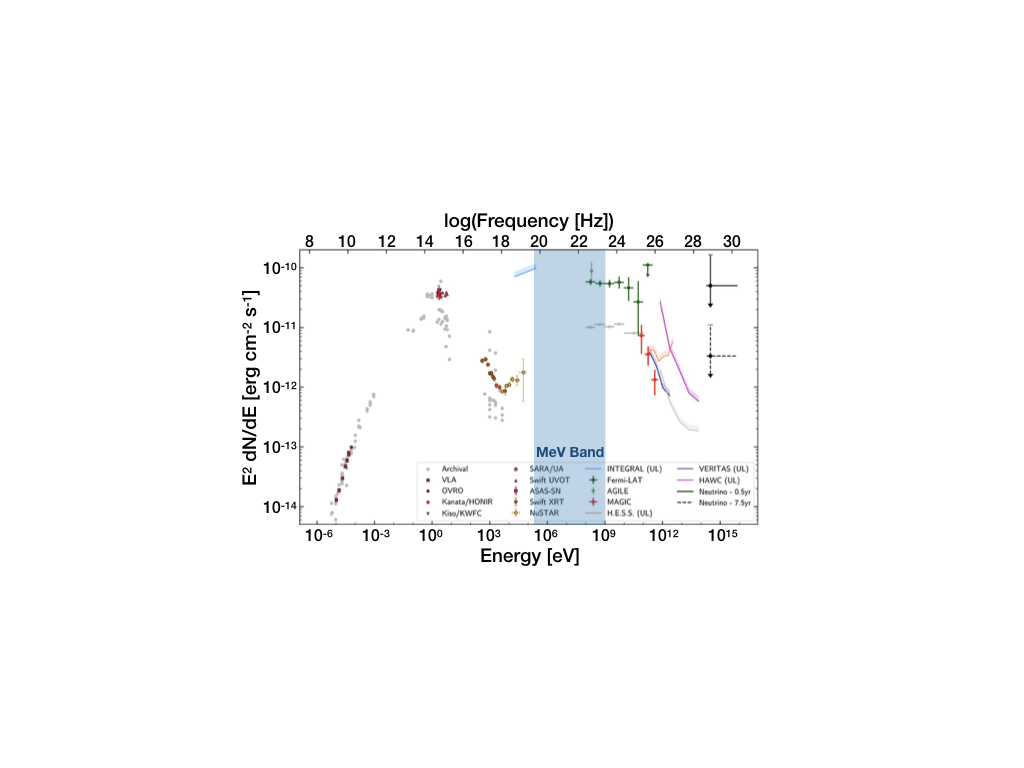}}
\vspace{-1.5ex}
\caption{\textit{\small Multi-messenger spectrum of TXS 0506+056\cite{2018Sci...361.1378I}. Future MeV measurements will play a crucial role in characterizing AGN emission processes and determining jet particle content.} \label{fig:txs0506p056_spec}}
\vspace{-5.ex}
\end{center}
\end{wrapfigure}
to high energies and possibly even to ultra-high energies.
High-energy multi-messenger observations of photons and neutrinos from AGN jets provide the means for identifying and characterizing their energetic particle populations, without which a more complete understanding of many aspects of jet phenomena would remain elusive. For instance, while the rotational energy of the SMBH \cite{blazna77} and/or the energy released through accretion \cite{1982MNRAS.199..883B} are often invoked to power the jet, the super-Eddington jet powers required by models that produce $\gamma$-ray emission through hadronic interactions 
challenge our notions of jet energetics and formation \cite{2015MNRAS.450L..21Z}. Radio synchrotron emission guarantees the presence of strong magnetic fields that likely drive and collimate the jet, but both radio and $\gamma$-ray emission are needed for measuring the energy content of the jet's electron population and determining the energy content in the magnetic fields. For jets that are initially dominated by magnetic fields as predicted by typical jet formation scenarios, the presence of energetic particles requires a mechanism for converting magnetic energy into bulk plasma kinetic energy and supplying energy to the particle populations through particle acceleration. Finally, energetic particles from AGN jets and outflows and the
radiation associated with their interactions provide a means for regulating star formation and black hole growth and heating the intracluster medium \cite{2017NatAs...1E.165H}, a process that has proved crucial to our understanding of the cosmological evolution of galaxies \cite{2013seg..book..555S}. Thus, high-energy multi-messenger observations of AGN jets in the coming decade will prove indispensable in understanding the phenomenology of their energetic particle populations.

At high energies, AGN jet emission can be explained by either Inverse Compton scattering of lower energy photons by accelerated electrons (purely leptonic scenarios) or by the radiative processes of accelerated protons or secondaries produced by inelastic collisions of accelerated protons with photons or other protons (lepto-hadronic scenarios). Of these scenarios, only those that include proton inelastic collisions are capable of producing neutrinos. Thus, the recent detection of a neutrino event coincident with a $\gamma$-ray flare in the blazar TXS 0506+056 \cite{2018Sci...361.1378I} could indicate the occurrence of proton inelastic collisions, though the associated radiation proved unable to fit the multiwavelength spectrum, overproducing the hard X-ray dip between the lower-energy hump and the higher-energy hump (see Fig. \ref{fig:txs0506p056_spec}). As such, future hard X-ray measurements of the dip will be crucial in determining jet particle content. We will also need MeV measurements to more fully characterize the high-energy hump of AGN spectra and distinguish among the various emission scenarios, particularly for those AGN in which the high-energy emission peaks in the MeV band. Going to the MeV band also provides the opportunity to measure MeV polarization in AGNs, a potentially powerful diagnostic tool as hadronic scenarios are expected to exhibit higher levels of polarization than leptonic scenarios \cite{2013ApJ...774...18Z}. A large collecting area, excellent time resolution, a wide field of view will enable variability studies that will also distinguish between leptonic and hadronic scenarios as the acceleration and cooling timescales for electrons and protons are different. Multiwavelength polarization campaigns (including X-ray and $\gamma$-ray spectropolarimetry), particularly in conjunction with the variability studies at high energies, will provide much needed insight into magnetic field structures and/or turbulent plasma cells in the jet and particle acceleration processes \cite{2014ApJ...780...87M,2016MNRAS.463.3365A,2018MNRAS.480.2872T}. It is worth noting that in leptonic-hadronic scenarios, these multi-messenger studies will be able to determine the energies for the proton population and assess whether AGN jets accelerate protons to ultra-high energies. On the other hand, even in the purely leptonic scenarios the MeV band will elucidate the composition of the jet -- MeV measurements of the low-energy cutoff in the spectra of high-frequency--peaked blazars will allow for determination of the electron energy density, which in comparison with the magnetic energy density, will determine whether these jets are dominated by pairs or by magnetic fields \cite{2018JHEAp..19....1D}.

\vspace{-3.0ex}
\section*{Gamma-ray Bursts}
\vspace{-2.0ex}

\begin{wrapfigure}{r}{0.6\textwidth}
\begin{center}
\vspace{-3.ex}
\resizebox{4.in}{!}
{\includegraphics[trim = 70mm 58mm 70mm 59mm, clip]{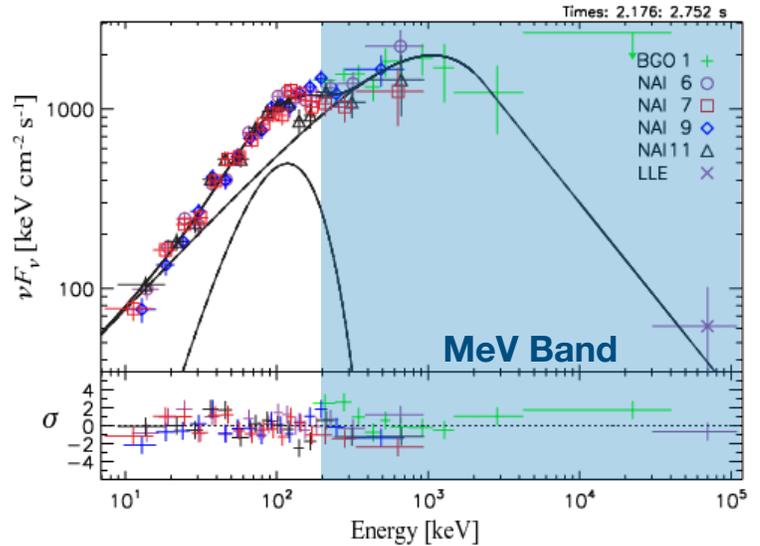}}
\vspace{-3.5ex}
\caption{\textit{\small GRB110721A spectrum\cite{Iyyani:2013yxa}. MeV measurements will play a crucial role in characterizing the GRB prompt emission.} \label{fig:grbspec}}
\vspace{-5.ex}
\end{center}
\end{wrapfigure}
Gamma Ray Bursts (GRBs) pinpoint the births of highly-magnetized neutron stars and/or stellar-mass black holes via either the collapse of hyper-massive stars or the merger of two neutron stars or a neutron star with a black hole. While there is, as yet, no universally accepted theoretical model that adequately explains every aspect of GRB observations, a useful framework for describing key features of the phenomena has emerged in the form of the fireball model \cite{1986ApJ...308L..43P,1986ApJ...308L..47G,1990ApJ...365L..55S,1993ApJ...405..278M}. In this model, the creation of the new compact object results in the release of a large quantity of gravitational energy on short timescales and within a very compact volume \cite{2004RvMP...76.1143P,Meszaros:2006rc,Peer:2015eek}. Much of this energy is carried away by neutrinos and gravitational waves; however, some amount of energy ($\gtrsim 10^{50}$--$10^{52}$ ergs) is released as a high-temperature, opaque fireball \cite{1978MNRAS.183..359C} consisting of $\gamma$-rays, $e^{\pm}$ pairs, and baryons that is funneled through a bipolar jet and expands at relativistic speeds. As the fireball expands, its optical depth decreases until the radiation can escape at the photospheric radius as thermal emission. On the other hand, much of the energy available to the fireball will be devoted to providing kinetic energy for any baryons present in the flow rather than to powering the thermal emission. In fact, for a high enough baryon load, not much energy is retained by the thermal radiation. Such a scenario requires some mechanism that will convert the kinetic energy of the baryons into radiation in order for an observable GRB to occur. Instabilities in the outflow can produce internal shocks that dissipate the kinetic energy \cite{1994ApJ...430L..93R} and accelerate particles via diffusive shock acceleration. In turn, the accelerated particles radiate via their radiative cooling processes (e.g., synchrotron and/or Inverse Compton in the case of electrons) giving rise to the non-thermal spectra that are commonly observed in the GRB prompt emission \cite{1993ApJ...413..281B} in the keV-MeV energy range. Further upstream, as the outflow plows into the surrounding medium, an external shock is formed that will also accelerate particles that radiate, producing the GRB afterglow observed at radio wavelengths up to X-rays and beyond.

Even though the fireball model has served as the guiding framework for theory for more than thirty years and certain key features likely remain valid, the model is faced with many unanswered questions and the physics of the GRB prompt emission is far from settled, particularly in light of recent observations by the \textit{Neil Gehrels Swift Observatory}\cite{Gehrels:2004aa} and \textit{Fermi}-GBM\cite{2009ApJ...702..791M} and \textit{AGILE}\cite{Tavani:2008sp}. For starters, the energy dissipation achievable through internal shocks is relatively inefficient unless the different parts of the ejecta have very different Lorentz factors \cite{1995Ap&SS.231..441M,Meszaros:2006rc}. Furthermore, the baryon load has to be finely tuned. Too many baryons would result in a sub-relativistic flow and either a very weak or a non-existent GRB. In fact, for bulk Lorentz factors $\Gamma \sim 100$, the mass of the baryons can be no more than $\sim 10^{-4}\ M_{\odot}$, which is difficult to explain with core-collapse or NS-NS/NS-BH merger progenitors of GRBs\cite{2004ASSL..302.....F}. On the other hand, lower baryon loads result in GRB spectra with a blackbody component associated with the photospheric emission. Recently, it has been demonstrated that several GRB spectra can be well fitted by including a blackbody component\cite{2009ApJ...706L.138A,2010ApJ...709L.172R,2011ApJ...727L..33G,2012ApJ...757L..31A,2013ApJ...770...32G,Iyyani:2013yxa,Ghirlanda:2013ara} (see Fig. \ref{fig:grbspec}). Including a blackbody component alters the parameters of the fitted nonthermal spectrum \cite{2013ApJ...770...32G}, particularly the low-energy spectral index ($\alpha$ in the empirical Band function\cite{1993ApJ...413..281B}), allowing it to be consistent with synchrotron emission at least in the regime in which electrons cool slowly\cite{1998ApJ...506L..23P,Burgess:2013kna,Yu:2014fsa}  (\textit{i.e.}, not violating the famous ``synchrotron line-of-death''\cite{1998ApJ...506L..23P} in the slow cooling regime). While these results are promising for the synchrotron interpretation of the non-thermal component, it is worth noting that in order for the electrons to cool slowly, the magnetic field would have to be relatively weak\cite{Zhang:2015bsa}. It is entirely possible that the magnetic fields are strong and could even dominate the jet dynamics, particularly in light of the commonly invoked mechanisms for extracting energy from a rotating compact object \cite{blazna77,1982MNRAS.199..883B}. In the presence of such strong magnetic fields, diffusive shock acceleration is not viable. For these latter reasons and in response to the aforementioned caveats of the fireball model, many in the field have turned to GRB models that include Poynting-flux--dominated outflows\cite{Meszaros:1996ww,Levinson:2006cw,Lyutikov:2005da,Giannios:2007yj,Tchekhovskoy:2008gq,Komissarov:2008ic,2011MNRAS.413.2031M,2011ApJ...726...90Z,2012MNRAS.419..573M,Sironi:2015eoa} as the energy dissipation occurs via magnetic reconnection and is expected to be more efficient\cite{Drenkhahn:2001ue,2002A&A...391.1141D,Komissarov:2008ic,2010MNRAS.402..353L,2010NewA...15..749T,2012MNRAS.419..573M,Sironi:2015eoa} and a finely-tuned baryon fraction is not required\cite{2004ASPC..312..449L}. As such, the nature of the GRB prompt emission remains the subject of considerable debate. 

MeV observations in the next decade will be crucial to disentangling the complex physical processes at work in the GRB prompt emission. We will need MeV measurements to more fully characterize GRB prompt emission spectra and obtain high fidelity measurements of non-thermal emission in GRBs -- not only does the measured spectrum peak in the MeV band in most cases (see Fig. \ref{fig:grbspec}), but in many cases ($\sim 25$\%\cite{2018JHEAp..19....1D}), the high-energy part of the spectrum (the $\beta$ index in the Band function) is poorly constrained. The MeV band is also where we expect to find cutoffs in spectrum\cite{2018ApJ...864..163V} related to the maximum energy of the emitting particle population and signals the end of efficient acceleration. The energy of the cutoff also depends on the bulk Lorentz factor; hence, MeV measurements of the cutoff will determine the bulk Lorentz factor\cite{2018JHEAp..19....1D}. An MeV telescope with an energy range that extends down to the hundreds of keV will also have continuity with instruments that operate in the tens to the hundreds of keV where the photospheric emission is expected to turn up. The connection with lower-energy instruments will allow for simultaneous fits of the multiband spectrum that will confirm the detection of the photospheric emission, if present, or determine if other types models (\textit{i.e.}, double broken power law) will be needed to explain the data.
To this end, we will also need time-resolved spectral measurements for the GRB prompt emission to fully confirm the presence of the photospheric emission as spectral evolution over the duration of the burst could mask the thermal component arising at different times result in time-integrated spectra that appears non-thermal\cite{Ghirlanda:2007yf}. Comparisons between the thermal component and the non-thermal component in prompt GRB emission will determine the baryon fraction and provide indications of the magnetization of the jet. Additionally, we will obtain further information on the jet magnetization by comparing the energy in the prompt emission to the energy in the afterglow emission\cite{Peer:2016xis}, leveraging the natural multiwavelength synergy of GRB observations. We will also be able to access the jet magnetization through MeV polarization measurements as synchrotron is highly polarized. An MeV telescope with a large effective area will be needed in order to achieve both the time resolution and the polarization sensitivity to conduct these measurements, as well as to provide population statistics by catching a large number of GRBs in the band where the energy output of many of them peak. Finally, MeV observations of GRBs provide an excellent opportunity for multi-messenger observations in conjunction with next-generation high-energy and very-high energy $\gamma$-ray instruments and neutrino telescopes. The recently announced \textit{MAGIC} detection of GRB190114C\cite{2019ATel12538....1M} indicates that GRBs are capable of producing $\gamma$-rays at energies $\sim$ hundreds of GeV and possibly even beyond. As such, GRB science presents an alluring opportunity for next-generation very-high energy telescopes, such as \textit{CTA}. It is also worth noting that even though there are expected to be only a handful of baryons, they could also be accelerated via the same mechanisms (diffusive shock acceleration or magnetic reconnection) that accelerate electrons and positrons. The photospheric emission and the radiative processes of the pairs will provide a target photon field with which the accelerated baryons will interact through the p$\gamma$ process. The pions produced through this process will not only supply more pairs to the jet, but will also produce $\gamma$-rays and neutrinos at GeV energies and beyond\cite{Waxman:1997ti,Waxman:1998yy,Murase:2008sp,Hummer:2011ms,Meszaros:2015krr}. All-sky searches for neutrinos from GRBs conducted by IceCube have already placed constraints on neutrino and ultra-high energy cosmic ray production in GRB fireballs\cite{Aartsen:2016qcr}. Next-generation neutrino telescopes, such as \textit{IceCube-Gen 2}, will probe even deeper into the parameter space. The recent observation of the merger of two neutron stars (GRB170817A/GW170817\cite{Monitor:2017mdv}) also speaks to the multimessenger synergy with currently-operating and next-generation gravitational wave detectors.

\vspace{-2.0ex}

\section*{The Promise of the Next Decade}

\vspace{-2.0ex}

The 2020s will usher in a new era of MeV astronomy that will revolutionize our quest to understand the most energetic particles and phenomena in our universe (for more details and many more examples, see \textit{e.g.}, the \textit{e-ASTROGAM} White Book \cite{2018JHEAp..19....1D}). In order to make the measurements necessary for both AGN and GRB jets, we need an MeV telescope with excellent continuum sensitivity. As GRBs are transient and AGNs exhibit flaring behavior, we need an MeV telescope with a large effective area, a very wide field of view enabling observations in survey mode, and the capability to quickly respond to targets of opportunity. In order to distinguish hadronic from leptonic emission and measure magnetization, we will also need a telescope that is capable of measuring polarization at MeV energies. Possible MeV missions include: \textit{AdEPT}\cite{2014APh....59...18H}, \textit{AMEGO}\cite{amego_2018_aa}, \textit{e-ASTROGAM}\cite{2018JHEAp..19....1D}, \textit{COSI}\cite{Kierans:2017bmv}, and \textit{SMILE}\cite{2015ApJ...810...28T}. This science will also leverage synergies with instruments currently operating in the MeV and GeV bands, such as \textit{Fermi}, \textit{AGILE}, and \textit{INTEGRAL}, next-generation optical survey telescopes (\textit{e.g.}, \textit{LSST}) and optical polarimeters, next-generation hard X-ray telescopes, such as \textit{HEX-P} or \textit{FORCE}, next-generation TeV telescopes, such as \textit{CTA}, and next-generation neutrino observatories, such as \textit{IceCube-Gen 2}.

\bibliography{cosacc}

\end{document}